# Sensitivity analysis for multivariable missing data using multiple imputation: a tutorial

Cattram D. Nguyen[1,2]*, Katherine J. Lee[1,2], Ian R. White[3], Stef van Buuren[4,5], Margarita Moreno-Betancur[1,2]

[1]Clinical Epidemiology and Biostatistics Unit, Department of Paediatrics, The University of Melbourne, Parkville, Victoria, Australia

[2]Clinical Epidemiology and Biostatistics Unit, Murdoch Children's Research Institute, Parkville, Victoria, Australia

[3]MRC Clinical Trials Unit at UCL, London, UK

[4]Dept of Child Health, Netherlands Organisation for Applied Scientific Research TNO, Leiden, The Netherlands

[5]Dept of Methodology and Statistics, University of Utrecht, Utrecht, The Netherlands

* Corresponding author: cattram.nguyen1@unimelb.edu.au

**Number of tables:** 2

**Number of figures:** 3

**ABSTRACT**

Multiple imputation is a popular method for handling missing data, with fully conditional specification (FCS) being one of the predominant imputation approaches for multivariable missingness. Unbiased estimation with standard implementations of multiple imputation depends on assumptions concerning the missingness mechanism (e.g. that data are "missing at random"). The plausibility of these assumptions can only be assessed using subject-matter knowledge, and not data alone. It is therefore important to perform sensitivity analyses to explore the robustness of results to violations of these assumptions (e.g. if the data are in fact "missing not at random"). In this tutorial, we provide a roadmap for conducting sensitivity analysis using the Not at Random Fully Conditional Specification (NARFCS) procedure for multivariate imputation. Using a case study from the Longitudinal Study of Australian Children, we work through the steps involved, from assessing the need to perform the sensitivity analysis, and specifying the NARFCS models and sensitivity parameters, through to implementing NARFCS using FCS procedures in R and Stata.

**INTRODUCTION**

Multiple imputation (MI) is a method for handling missing data that is widely used in medical and epidemiological research.[1] MI consists of two stages. Firstly, the missing values are replaced with multiple imputed values drawn from a predictive distribution estimated from the observed data, resulting in multiple completed datasets. In the second stage, the analysis of interest is applied to each completed dataset, resulting in multiple estimates of the target quantity. The multiple estimates are then combined using "Rubin's rules" to obtain a single overall estimate of the parameter of interest, along with a variance estimate that reflects both within- and between-imputation uncertainty.[2,3]

There are two main approaches for imputing missing data in multiple variables: joint modelling and fully conditional specification (FCS). The joint modelling approach models incomplete variables using a joint multivariate distribution, and Markov Chain Monte Carlo methods are used to fit the model and draw imputed values.[4] The FCS approach, also known as multivariate imputation by chained equations (MICE), involves specifying univariate imputation models for each incomplete variable, conditional on other variables.[5] Generalised linear regression models are often used, tailored to the variable type, such as logistic regression for imputing binary variables. In this paper, we focus on the FCS approach, which is widely used due to its flexibility and availability in mainstream statistical packages[6,7].

As with all incomplete data analyses, when using MI, the user needs to make assumptions about the missing data, some of which are untestable. Rubin proposed a framework for describing the possible missingness mechanisms.[8] With one incomplete variable, data are considered to be "missing at random" (MAR) if the probability of missingness does not depend on the missing values given observed values of other variables, and "missing not at random" (MNAR) otherwise. A special, often unrealistic, case of MAR is when the probability of missingness does not depend on the values of complete or incomplete variables, in which case the data are said to be "missing completely at random" (MCAR). In the context of multivariable missingness, the formal definitions of the missingness mechanisms have not been widely understood. Seaman et al.[9] clarified that the assumption that is required to guarantee valid frequentist inference with likelihood-based methods, including standard implementations of MI, is the so-called "everywhere MAR" assumption. This holds if, across repeated samples, the probability of a given pattern of missingness across the incomplete variables depends on the observed but not the unobserved values in that pattern. Given our focus on the multivariable missingness context, we will henceforth use the term MAR to mean everywhere MAR, and MNAR if the data are not everywhere MAR. In many cases, the MAR assumption is likely to be unrealistic, but it is not possible to examine whether it holds without knowing the values of the missing data. Instead, external, subject-matter knowledge is required to examine the plausibility of this assumption. It is therefore important to conduct sensitivity analyses to assess the sensitivity of MI inferences to departures from MAR when these departures are deemed plausible.

Unfortunately, in the multivariable missingness context, examining the (im)plausibility of MAR using subject-matter knowledge is extremely difficult. Furthermore, although MAR is sufficient for MI to be unbiased, it is not necessary. Whether MI is unbiased under an MNAR mechanism depends on the target parameter and specific form of MNAR missingness mechanism. An alternative framework[10-12] uses directed acyclic graphs (DAGs) to depict missingness assumptions, which facilitates their assessment in practice using subject-matter knowledge. Additionally, this framework has elucidated which MNAR missingness mechanisms warrant sensitivity analyses, for example when a variable causes its own missingness. Specifically, recent work[12,13] determined typical MNAR missingness mechanisms where common target parameters of interest are not identifiable or "recoverable", which means that these parameters cannot be consistently estimated using the observable data alone. When such departures from MAR are deemed plausible, then sensitivity analyses to examine the impact of such mechanisms on findings are warranted. Meanwhile, with other typical MNAR mechanisms, target parameters are still recoverable, rendering sensitivity analyses unnecessary.

Sensitivity analyses to departures from MAR are increasingly recommended in guidelines for handling missing data;[14,15] however, the uptake of such sensitivity analyses is not widespread.[1,16,17] In the missing data literature, two general frameworks have been proposed for conducting sensitivity analyses to departures from MAR, primarily in the context of one incomplete variable or incomplete repeated measures: the selection model and pattern-mixture modelling approaches.[18-21] These two approaches are based on different factorisations of the joint model for the data and the missingness mechanisms.[18] In the context of one incomplete variable, the delta-adjustment method, within the pattern-mixture modelling framework, involves regression modelling of the incomplete variable conditional on its missingness indicator.[22] The regression coefficients associated with the missingness indicator, referred to as "deltas", are not estimable from the data and are thus also referred to as sensitivity parameters that are set to user-specified values. This approach can be implemented within MI, where departures from an MI analysis under MAR are achieved by shifting imputations by the specified deltas[5]. The not at random fully conditional specification (NARFCS), the focus of this paper, is an extension of the delta-adjustment method to the setting with multiple incomplete variables within the FCS framework.[23,24] Applications of the NARFCS procedure have been described in the MI literature,[5,25] and it can now be implemented in R and Stata software.[7,26,27] Given that multivariable missingness is common, that FCS is in wide use and NARFCS extensions are accessible in mainstream software, NARFCS provides a convenient method for carrying out sensitivity analyses, which is why we focus on it. However, there is little guidance for researchers on how to use this relatively recent approach.

The aim of this paper is to fill this gap by providing a tutorial on how to conduct sensitivity analyses using NARFCS, with worked examples and syntax in R and Stata. In the next section we introduce the motivating example from the Longitudinal Study of Australian Children. This is followed by sections covering important preliminary background: overviews of FCS, the delta-adjustment method and NARFCS, and an illustration of FCS applied to the motivating example. We then introduce and take readers through a roadmap for carrying out a sensitivity analysis using NARFCS, which includes the following steps: i) assessing the need to carry out a sensitivity analysis, ii) specifying the delta-adjusted models, iii) assigning values to the sensitivity parameters, iv) conducting NARFCS and v) reporting results. These steps are also illustrated using the motivating example.

**MOTIVATING EXAMPLE: THE LONGITUDINAL STUDY OF AUSTRALIAN CHILDREN (LSAC)**

The Longitudinal Study of Australian Children (LSAC) is a nationally representative, multi-wave cohort study of childhood development in Australia.[28] The motivating example involves a simplified version of an investigation using LSAC data described in Bayer et al.[29] that has also been considered in previous methodological papers.[12,30] The case study uses data from 4882 children in the LSAC Kindergarten cohort to examine the association between maternal mental illness in early childhood (wave 1, 4-5 years) and later child behaviour (wave 3, 8-9 years). The outcome variable is the child's score on the Strengths and Difficulties Questionnaire (SDQ) at wave 3, where higher scores indicated increased behavioural difficulties.[31] The exposure variable is a binary indicator of probable serious mental illness at wave 1, defined as an average score less than 4 on the Kessler-6 Psychological Distress Scale.[32] There are a number of measured potential confounders of this relationship, all measured at wave 1: sex of the child, whether the child has a sibling, maternal high school completion, maternal age, consistent parenting, financial hardship, child behaviour (SDQ score at wave 1), maternal smoking, maternal risky alcohol drinking and child's physical functioning score. The target analysis is a multivariable linear regression model with mean specification given by:

$$E(sdqw3) = \beta_0 + \beta_1 matmhw1 + \beta_2 sex + \beta_3 siblings + \beta_4 matedu + \beta_5 matage + \beta_6 conspar + \beta_7 finhard + \beta_8 basesdq + \beta_9 matsmok + \beta_{10} matalc + \beta_{11} physfunc$$

where variable names are defined in Table 1. The parameter of interest is $\beta_1$, the regression-adjusted difference in SDQ scores between children with and without mothers with probable serious mental illness. Under certain assumptions, this can be interpreted an estimate of the causal effect of maternal mental illness on childhood behaviour.[33]

**[Table 1 about here]**

As shown in Table 1, 1142 (23%) have missing outcome data and 738 (15%) have missing exposure data. There are also missing values for three confounders: maternal smoking (n=771, 16%), maternal risk alcohol drinking (n=885, 18%) and child's physical functioning score (n=742, 15%). Only 3245/4882 (66%) have complete data for all twelve variables in the target analysis.

**BACKGROUND ON FCS AND NARFCS**

In this section, we introduce the FCS procedure for imputing multiple incomplete variables. This is followed by a description of the delta-adjustment method for sensitivity analyses in the context of one incomplete variable. We then describe the NARFCS procedure, which is an extension of FCS implementing the delta-adjustment method for the setting with multiple incomplete variables.

*Fully conditional specification (FCS)*

FCS is a multivariate imputation method that imputes data iteratively on a variable-by-variable basis using a series of univariate imputation models, one for each incomplete variable.[34] To define this formally, we introduce some notation, considering for simplicity in this background section (in contrast to the motivating example) a setting with only two confounders, one complete and one incomplete. Let $Y$ denote an incomplete outcome variable, $X$ an incomplete exposure variable, and $Z_2$ the incomplete confounder. $\boldsymbol{M} = (M_Y, M_X, M_{Z_2})$ are the corresponding missingness indicators, with $M_Y = 1$ if $Y$ is missing and otherwise $M_Y = 0$; $M_X = 1$ if $X$ is missing and otherwise $M_X = 0$; and $M_{Z_2} = 1$ if $Z_2$ is missing and otherwise $M_{Z_2} = 0$. $Z_1$ is the completely observed confounder.

The FCS procedure involves specifying a series of models, one for the conditional distribution of each incomplete variable given the other variables. For example, suppose $Y$ is continuous, and $X$ and $Z_2$ are binary, then the FCS models could be:

$E(Y|Z_1, Z_2, X) = \alpha_0 + \alpha_1 Z_1 + \alpha_2 Z_2 + \alpha_3 X$

$\text{logit}\{P(X = 1|Z_1, Z_2, Y)\} = \gamma_0 + \gamma_1 Z_1 + \gamma_2 Z_2 + \gamma_3 Y$

$\text{logit}\{P(Z_2 = 1|Z_1, Y, X)\} = \zeta_0 + \zeta_1 Z_1 + \zeta_2 Y + \zeta_3 X$

The FCS procedure begins by randomly drawing imputed values for the missing data from the observed data. Each variable is then re-imputed in turn, by performing draws from the specified univariate conditional model fitted to the observed data and the most recent imputed values for other variables. One iteration consists of cycling through and imputing each of the incomplete variables in turn, with several such iterations conducted until convergence to obtain one imputed dataset.[34,35] The algorithm is repeated $m$ times to produce $m$ imputed datasets. Target analyses are applied to each of the $m$ imputed datasets and results are combined using Rubin's rules.[3]

*Delta-adjustment method for one incomplete variable*

The delta-adjustment method is a sensitivity analysis approach within the pattern-mixture modelling framework. Within this method, the incomplete variable is imputed using a regression model that is conditional on its own missingness indicator, i.e. it allows for distinct models for those with and without complete data, reflecting a pattern-mixture model. This encodes an MNAR mechanism under which sensitivity to departures from MAR can be assessed.[3] For instance, if in the example only $Y$ were incomplete (so that $Z_1$, $Z_2$ and $X$ are all complete), then an example of a pattern-mixture model for this continuous variable is:

$E(Y|Z_1, M_Y) = \theta_0 + \theta_1 Z_1 + \theta_2 Z_2 + \theta_3 X + \delta M_Y$

The $\delta$ coefficient (i.e. delta, also known as the sensitivity parameter) represents the difference between non-respondents and respondents in the conditional mean of $Y$ given $Z_1$, $Z_2$ and $X$. This parameter is not identifiable (i.e. not estimable) from the data and must be assigned a value by the user (more detail on how this might be done is provided in subsequent sections).

Given values of the sensitivity parameter(s), the delta-adjustment method can be easily implemented within an MI procedure in practice. For a continuous incomplete variable, missing values are first multiply imputed assuming MAR using standard methods. To incorporate an assumed MNAR mechanism as in the above model, the chosen $\delta$ value is added as a fixed quantity to the imputed values. The multiple imputed datasets are then analysed using standard approaches and results are combined using Rubin's rules. A similar approach can be used for binary variables, but $\delta$ is incorporated as an offset in the univariate imputation model e.g. in a logistic regression model. In this case, $\delta$ represents a shift in the log-odds of $Y$ between non-respondents and respondents.[22,30]

There have been several applications of the delta-adjustment method in the missing data literature.[22,25,36-38] Usually, a range of values for $\delta$ is selected and the procedure repeated for each of the $\delta$ values[25]. More complex models can be considered for the dependency of the incomplete variable on its missingness indicator, for example, in addition to an intercept shift, it is also possible for there to be interaction terms with other variables (e.g. trial arm in a randomised controlled trial), so that there are multiple sensitivity parameters/deltas.[25,39]

*Not at random fully conditional specification (NARFCS)*

NARFCS[23] extends the delta-adjustment method to the setting with multiple incomplete variables within the FCS framework. In brief, like standard FCS, a univariate regression model is posited for each incomplete variable, but with NARFCS, the regression model for an incomplete variable can include its own missingness indicator as a predictor, with the corresponding regression coefficient being a sensitivity parameter that represent a shift in the distribution of missing values. Specifically, using notation as above, for each incomplete variable we posit a model for the conditional distribution given other variables and, in some cases, its own missingness indicator:

$$E(Y|Z_1, Z_2, X, M_Y) = \eta_0 + \eta_1 Z_1 + \eta_2 Z_2 + \eta_3 X + \delta_Y M_Y$$

$$\text{logit}\{P(X = 1|Z_1, Z_2, Y, M_X)\} = \lambda_0 + \lambda_1 Z_1 + \lambda_2 Z_2 + \lambda_3 Y + \delta_X M_X$$

$$\text{logit}\{P(Z_2 = 1|Z_1, Y, X, M_{Z_2})\} = \xi_0 + \xi_1 Z_1 + \xi_2 Y + \xi_3 X.$$

Here again, the sensitivity parameters ($\delta_Y$ and $\delta_X$ in this example) must be assigned a value by the user (more details below) as they cannot be estimated from the data. Using these models, MI is performed following the same iterative algorithm as described above for FCS to obtain imputed values that incorporate departures from MAR. Example applications of this method are described in Tompsett et al.[23] and Van Buuren et al.[5]

**APPLYING FCS TO THE MOTIVATING EXAMPLE**

We now illustrate how to carry out MI within FCS under MAR in the LSAC example, which is important groundwork for how NARFCS is then implemented. Throughout the tutorial, we focus on implementation using the "mice" package in R, with equivalent code in Stata provided in the Supplement. To enable readers to use the code, in the Supplement we have also provided a dataset that was simulated (loosely) based on the design and associations in the real LSAC data (note that readers will not be able to reproduce the results in the paper as these are based on the real data that we cannot freely share).

We begin by loading the "mice" package as well as the data in R:

```
# Load the mice package
library(mice)

# Read the data
load("datmis.Rda")
```

The dataset contains twelve variables. Inspection of the data confirms that there are missing values in the outcome *sdwq3*, exposure *matmhw1* and three confounders *matsmok*, *matalc* and *physfunc*.

```
#Obtain summary of dataset and inspect missing values per variable
summary(datmis)
 sex              siblings         matedu           matage
 Min.   :0.0000   Min.   :0.0000   Min.   :0.0000   Min.   :19.83
 1st Qu.:0.0000   1st Qu.:1.0000   1st Qu.:0.0000   1st Qu.:31.75
 Median :1.0000   Median :1.0000   Median :1.0000   Median :35.09
 Mean   :0.5078   Mean   :0.8882   Mean   :0.5868   Mean   :35.15
 3rd Qu.:1.0000   3rd Qu.:1.0000   3rd Qu.:1.0000   3rd Qu.:38.75
 Max.   :1.0000   Max.   :1.0000   Max.   :1.0000   Max.   :66.00

 conspar          finhard          basesdq          matsmok
 Min.   :1.200    Min.   :0.0000   Min.   : 0.000   0   :3165
 1st Qu.:3.600    1st Qu.:0.0000   1st Qu.: 5.000   1   : 946
 Median :4.200    Median :0.0000   Median : 9.000   NA's: 771
 Mean   :4.049    Mean   :0.5336   Mean   : 9.349
 3rd Qu.:4.600    3rd Qu.:1.0000   3rd Qu.:12.000
 Max.   :5.000    Max.   :6.0000   Max.   :35.000

 matalc       physfunc          matmhw1      sdqw3
 0   :3823    Min.   :  0.00    0   :3282    Min.   : 0.000
 1   : 174    1st Qu.: 78.12    1   : 862    1st Qu.: 4.000
 NA's: 885    Median : 84.38    NA's: 738    Median : 6.000
              Mean   : 82.77                 Mean   : 7.478
              3rd Qu.: 90.62                 3rd Qu.:10.000
              Max.   :100.00                 Max.   :35.000
              NA's   :742                    NA's   :1142
```

We will now use `mice()` to impute the missing values. Within `mice()`, the predictors in each imputation model can be tailored using either the formula syntax or the predictor matrix syntax. We will use the `make.predictorMatrix()` function to create a predictor matrix, which is a square matrix of size equal to the number of variables in the dataset. The matrix consists of 0s and 1s, where a cell is assigned a value 1 if the column variable is to be included in the imputation model for the row variable; otherwise it is equal to 0. In `mice()`, by default, the univariate imputation model for each incomplete variable will include all other variables (therefore this step was not essential, but we have included it to be explicit).

Within `mice()`, the `method` argument is a vector used to specify the method for imputing each variable in the order in which they appear in the dataset. We will use `norm` to impute continuous variables (*physfunc, sdqw3*) using linear regression, and `logreg` to impute binary variables (*matsmok*, *matalc*, *matmhw1*) using logistic regression. For completely observed variables that do not require imputation, an empty method is specified using `""`. Additional arguments include: `m=40` to generate 40 imputed datasets, `predictorMatrix = predMatrix` to specify the predictors to be included in the FCS imputation models, `maxit=10` to specify 10 iterations, `seed=3857814` to set a seed for reproducibility, and `print=F` to suppress the output. We chose 40 imputations based on the rule of thumb that the number of imputations should be at least equal to the percentage of incomplete cases (which was 34% for the LSAC example), and 10 iterations based on findings that low numbers of iterations (5-20) is often sufficient for convergence of the FCS algorithm.[35,40] The 40 imputed datasets are stored in the object `imp`.

```
# Create predictor matrix argument for the mice function
predMatrix<-make.predictorMatrix(datmis)

# Impute missing values using mice
imp <- mice(datmis, m=40, predictorMatrix=predMatrix,
method=c(rep("",7),"logreg", "logreg","norm", "logreg", "norm"), maxit=10,
seed=3857814, print=F)
```

At this stage, we should check the proposed imputation model. This could include, for example, summary statistics or visual displays (e.g. boxplots) of the observed and imputed data. These summaries and graphs can be used to check whether the imputed values are plausible, and whether any differences between the observed and imputed values are expected given likely missing data mechanisms. Convergence of the FCS algorithm can also be assessed using, for example, trace plots of parameters of interest against iteration number. For further information on imputation model checking and implementation of checks in R and Stata we refer readers to van Buuren et al.[41] and Nguyen et al.[42]

Proceeding with the proposed imputation model above, we now fit the target regression analysis to each of the imputed datasets and pool the results using Rubin's rules. The `with()` function is used to fit the linear regression analysis on each of the imputed datasets, and the `pool()` function combines the multiple parameter estimates. We extract the estimates of our parameter of interest, which is the regression coefficient of the exposure *matmhw1*.

```
# Fit analysis model to each dataset

fit <-
with(imp,lm(sdqw3~matmhw1+sex+siblings+matedu+matage+conspar+finhard+basesd
q+matsmok+matalc+physfunc))

# Pool the result and present the point estimate and 95% confidence
intervals

round(summary(pool(fit), conf.int = TRUE, conf.level = 0.95)[2,
c(2,7,8)],2)

estimate    2.5%        97.5%
0.65        0.22        1.08
```

Using FCS, the estimate of the regression coefficient for maternal probable mental illness is 0.65 (95% confidence interval: 0.22 -1.08). This suggests that maternal mental illness may be associated with slightly higher SDQ scores.

**ROADMAP FOR CONDUCTING A NARFCS SENSITIVITY ANALYSIS**

In this section, we provide a roadmap for carrying out a sensitivity analysis using NARFCS, illustrated using the LSAC case study. See Figure 1 for a schematic and summary of this roadmap. It is worth noting that pre-planning all analyses, whether primary or sensitivity analyses, is recommended.[43] It is common to plan for the primary analysis to be based on the MAR assumption, such as the analysis above using FCS, given this analysis does not need external information such as elicited values of sensitivity parameters. Then researchers need to examine whether they need to plan for a sensitivity analysis to departures from MAR. Although sensitivity analyses are increasingly recommended,[14] it is important to consider whether a sensitivity analysis is warranted for a specific analysis.

[Figure 1 about here]

*Step 1: Decide whether to conduct a sensitivity analysis*

To decide whether to carry out a sensitivity analyses we need to examine whether we expect departures from MAR to be plausible. This decision generally involves seeking input from the study team and content experts about the likely reasons underlying the missing data, which can be done through an informal or formal process. For example, our LSAC collaborators may believe that those with mental illness are less likely to return their questionnaires for reasons beyond other measured characteristics, implying a potential departure from MAR. Based on these informal discussions with our colleagues, we may decide that a sensitivity analysis would be helpful to understand the impact of this departure from MAR on study results.

A more formal approach to elicit content expert knowledge about the reasons for missing data may be to construct a directed acyclic graphs (DAGs) in discussion with the experts. Specifically, interest here is on missingness-DAGs (m-DAGs or m-graphs), which are extensions to causal diagrams that include variable-specific missingness indicators and thus enable assumptions about the relationships between variables and missingness in individual variables to be depicted.[10-12,44] This graphical approach is helpful in cases with multivariable missingness where the details of the missing data mechanisms are complex to describe by other means. Once an m-DAG is posited in discussion with content experts, it can be used to formally guide decisions about whether a sensitivity analysis is required, based on whether target parameter is likely to be "recoverable" (i.e. non-parametrically identifiable) given the depicted causal structure. If the parameter is recoverable, then a sensitivity analysis is not warranted. But if the parameter is not recoverable, then it means that it cannot be consistently estimated from the patterns and associations in the observed data, e.g. using standard FCS, without needing to invoke external information on the extent to which missing values may differ from observed ones. Therefore a sensitivity analysis incorporating external data about these differences (e.g. through sensitivity parameters/deltas) is required.

To illustrate this approach with the LSAC example, we first set some notation. For simplicity, we use $\boldsymbol{Z_1}$ and $\boldsymbol{Z_2}$ to represent vectors of complete and incomplete confounders respectively, with the incomplete confounders $\boldsymbol{Z_2}$ being *matsmok*, *matalc* and *physfunc* (cf.Table 1). As above, $M_X$ refers to the missingness indicator for the exposure $X$ (maternal mental illness), with $M_X = 1$ if $X$ is missing, otherwise $M_X = 0$. Similarly, $M_Y$ is the missingness indicator for the outcome variable $Y$ (child behaviour). $M_{Z_2}$ is the missingness indicator for $\boldsymbol{Z_2}$, with $M_{Z_2} = 1$ if any variables in $Z_2$ are missing, otherwise $M_{Z_2} = 0$. $U$ denotes the vector of unmeasured common causes of exposure and confounders, while $W$ is the vector of unmeasured common causes of the missingness indicators.

**[Figure 2 about here]**

Figure 2 shows one plausible m-DAG for the case study (based on the m-DAG in Figure 1 of Moreno-Betancur et al).[12] To develop this m-DAG, we needed to assess the likely relationships between incomplete variables and missing data indicators. As described by Moreno-Betancur et al. (Figure 3 of that paper), the arrow from exposure $X$ to its missingness indicator $M_X$ seems likely, as maternal mental health issues could increase the failure to return survey questionnaires.[12] The arrow between the incomplete confounders $Z_2$ to $M_{Z_2}$ also seems likely, for example, as there may be stigma associated with maternal drinking and alcohol use, leading to missing responses to these questions. The arrow from $Y$ to its missingness indicator $M_Y$ is also plausible as current child difficulties could lead to failure to return questionnaires.

Based on the assumptions depicted by this m-DAG, we examine the recoverability of the target parameter (regression adjusted exposure-outcome association) and decide whether a sensitivity analysis is needed. To do this, several approaches can be used.[10,11,13,43] Here we utilise existing recoverability results derived by Moreno-Betancur et al.[12]. Specifically, the authors outline m-DAGS representative of all possible missingness mechanisms in a point-exposure study like this one, including our LSAC m-DAG in Figure 2, and provide results about whether specific parameters are recoverable for each of these m-DAGs. Using these results, we can determine that our target parameter, the regression adjusted exposure-outcome association, is not recoverable. This indicates that a sensitivity analysis is required, and we decide to use NARFCS because there are multiple incomplete variables.

*Step 2: Specify the delta-adjusted models*

Having decided to carry out a sensitivity analysis using NARFCS, the next step is to specify the delta-adjusted models. As with usual FCS, a univariate model for each incomplete variable is specified, but we include the missingness indicators $M_Y$ and $M_X$ in the models for $Y$ and $X$, respectively, in line with the corresponding arrows depicted in the m-DAG (or with informal discussions with collaborators suggesting associations between exposure and outcome and their missingness). It would also be possible to include missingness indicators in the models for the incomplete confounders (e.g. as we considered the arrow between the $Z_2$ to $M_{Z_2}$ to be likely in the m-DAG), however, theoretical results in Moreno-Betancur et al.[12] indicate that in most cases these arrows or associations do not preclude consistent estimation of the regression coefficient, which is the key parameter of interest. Using the notation presented in Table 1, the NARFCS models for the case study are as follows:

$\text{logit}\{P(X = 1|Z_1, Z_2, Y, M_X)\} = \rho_0 + \rho'_1 Z_1 + \rho'_2 Z_2 + \rho_3 Y + \delta_X M_X$

$E(Y|Z_1, Z_2, X, M_Y) = \phi_0 + \phi'_1 Z_1 + \phi'_2 Z_2 + \phi_3 X + \delta_Y M_Y$

$\text{logit}\{P(Z_{21} = 1|Z_1, Z_{22}, Z_{23}, Y, X)\} = \chi_0 + \chi'_1 Z_1 + \chi_2 Z_{22} + \chi_3 Z_{23} + \chi_4 Y + \chi_5 X$

$\text{logit}\{P(Z_{22} = 1|Z_1, Z_{21}, Z_{23}, Y, X)\} = \tau_0 + \tau'_1 Z_1 + \tau_2 Z_{21} + \tau_3 Z_{23} + \tau_4 Y + \tau_5 X$

$$E(Z_{23}|\mathbf{Z_1}, Z_{21}, Z_{22}, Y, X) = \psi_0 + \boldsymbol{\psi'_1}\mathbf{Z_1} + \psi_2 Z_{21} + \psi_3 Z_{22} + \psi_4 Y + \psi_5 X$$

where $\mathbf{Z_2} = (Z_{21}, Z_{22}, Z_{23})$. The two sensitivity parameters, $\delta_Y$ and $\delta_X$, are the coefficients for the missingness indicators for $Y$ and $X$, respectively. $\delta_Y$ represents the difference in the conditional mean of the outcome between non-respondents and respondents given other variables in the imputation model; $\delta_X$ represents the difference in the conditional log-odds of exposure (i.e. maternal mental illness) between non-respondents and respondents given other variables in the imputation model (also described in the literature as the log informative missingness odds ratio).[45]

### Step 3: Assign values to the sensitivity parameters

The next step in conducting a NARFCS sensitivity analysis is to select values for the sensitivity parameters, as, by definition, these parameters cannot be estimated from the observed data. It is worth stressing that the model in Step 2 needs to have been chosen to make these parameters reasonably easy to interpret and thus elicit. Given the uncertainty about the values of these parameters, an approach is to specify a plausible range of values for each delta parameter and repeat the NARFCS procedure for several values within this range. This is the approach we consider for the LSAC example, where the ranges of values were based on elicited distributions (see below). An alternative approach is to specify or elicit distributions for the sensitivity parameters and then draw values for each delta parameter for each imputed dataset.[46]

The process of deciding on a plausible range or distribution of values for the delta parameters can be done informally or formally. An informal approach is to select, for each sensitivity parameter, a plausible range of values or distribution based on an informal discussion with content experts, or entirely without external input. One such approach is a tipping point analysis, where the sensitivity analysis is carried out across a range of values to assess whether there is a point at which qualitative conclusions change (or "tip" to a different conclusion).[20,23] Note however that this tipping point approach focuses on binary conclusions, i.e. whether MNAR departures are likely to change conclusions about rejecting or accepting null hypotheses, a viewpoint that is increasingly discouraged.[47]

A more formal approach for selecting sensitivity parameter value ranges or distributions is expert elicitation. This involves asking experts for (ranges of) plausible values based on their expectation regarding the differences between those with and without missing values in the distributions of the incomplete variables. Various formats can be used to carry out the elicitation, from email questionnaires and web applications to face-to-face interviews,[39,48,49] but an interactive process that allows feedback is considered important to ensure that the elicited information reflects the expert's views as they may be unfamiliar with the elicitation process.[50] The expert elicitation can be carried out on individuals or in groups. If conducting elicitations from multiple experts, the elicited values may need to be combined to obtain a pooled distribution for each sensitivity parameter.[49]

One challenge in the elicitation of NARFCS sensitivity parameters is that they are conditional in nature, i.e. they represent the shift in values between respondents and non-respondents conditional on all other variables in the imputation model. Tompsett et al.[23] refer to these parameters as conditional sensitivity parameters (CSPs) and distinguish them from marginal sensitivity parameters (MSPs). MSPs are typically easier to interpret and to elicit than CSPs, as it can be challenging to elicit conditional information from experts. For example, in the LSAC analysis, to provide values for $\delta_Y$, experts would need to understand the likely differences in mean between respondents and non-respondents who are matched on maternal mental illness and all of the complete and incomplete confounders. To overcome the difficulties in eliciting CSPs, Tompsett et al.[23] propose a calibration procedure, whereby MSPs are elicited from experts and these MSPs are converted or "calibrated" to obtain conditional values. For more detail on the calibration procedure, see Tompsett et al.[23]

For the LSAC example, we need to elicit likely values for $\delta_Y$ and $\delta_X$. For our NARFCS analysis, the values of the sensitivity parameters were based on those selected by Hayati-Rezvan et al., who performed expert elicitation for the same LSAC example.[30] In brief, the authors invited three experts in child health and development to participate in the elicitation task. Face-to-face interviews were conducted with each expert individually, guided by a written questionnaire. The experts provided expected minimum, lower quartile, median, upper quartile and maximum MSP values for both the exposure and outcome. These values were converted into individual probability density functions, and then combined to create a pooled probability distribution for each parameter. There were differences in the individual probability density functions across the three experts, with one expert's distribution being right skewed while distributions for the other two experts were fairly symmetrical (see Figure 1 in Hayati-Rezvan et al. [30]). Therefore, when conducting the sensitivity analyses, Hayati-Rezvan et al. [30] used selected quantiles of the pooled expert-elicited distributions (see Table 2). As the sensitivity parameters elicited by Hayati-Rezvan et al. were marginal, we used the procedure of Tompsett et al. to calibrate these to obtain CSPs to be used in the NARFCs procedure (see Supplementary material and Tompsett et al. for R code used for calibration).[23] Of note, in this example the experts were asked to provide information on the difference in proportion with maternal mental illness, as this quantity is easy to interpret; therefore the differences in proportions were then converted to log-odds ratios for the sensitivity analysis.

**[Table 2 about here]**

### Step 4. Conduct NARFCS sensitivity analyses

Once the delta-adjusted models and sensitivity parameter values have been selected, the NARFCS sensitivity analyses can be carried out. Within the "mice" package in R, this will be done using the `mice.impute.mnar.norm()` and `mice.impute.mnar.logreg()` functions, which impute data under delta-adjusted linear and logistic imputation models, respectively.[24,41] Specifically, for variables that are being imputed with delta-adjustment (i.e. *sdqw3* and *matmhw1*), we will use `mnar.norm` and `mnar.logreg` for the `method` argument, respectively. For variables being imputed without delta-adjustment, we will use `logreg` to impute the binary variables (*matsmok* and *matalc*), and `norm` to impute the continuous variable *physfun* (as done in the FCS section above).

To carry out NARFCS, the values of the sensitivity parameters must also be specified. This is done via the `blots` argument of mice(), which is a named list that is used to pass arguments to univariate imputation models. For each variable to be imputed under MNAR, the corresponding element in `blots` is a list with a required parameter `ums`. The `ums` specification must include a term corresponding to an intercept term. For example, in the syntax below we specify an intercept shift of $\delta_Y = 0.78$ in the imputation model for *sdqw3* and an intercept shift of $\delta_X = 0.11$ for *matmhw1* (these shifts correspond to the median values of the (elicited and calibrated) sensitivity parameters shown in Table 2). These values are stored in the object `mnar.blot`.

```
mnar.blot <- list(sdqw3 = list(ums = "0.78"), matmhw1 = list(ums = "0.11"))
```

Multiple terms can be included in `ums`, separated by either an "+" or "-" sign. For non-intercept terms, the sensitivity parameter value comes first, followed by the star symbol "*" and the predictor variables. For example, the `ums` specification could be "`0.78 + 0.2*matage`" if the specified univariate imputation model included an interaction between the missingness indicator and maternal age.

We are now ready to carry out NARFCS. In the line of code below, the sensitivity parameter values (stored in `mnar.blot`) have been passed to the `mice()` function using the `blots` argument. The output of `mice()` is stored in `impNARFCS`.

```
impNARFCS <- mice(datmis, m=40, predictorMatrix=predMatrix,
method=c(rep("",7),"logreg", "logreg","norm", "mnar.logreg",
"mnar.norm"), blots=mnar.blot, maxit=10, seed=23435, print=F)
```

As with standard FCS, we fit the linear regression analysis to each imputed dataset and pool the results using Rubin's rules:

```
fit <- pool(with(impNARFCS,
lm(sdqw3~matmhw1+sex+siblings+matedu+matage+conspar+finhard+ba
sesdq+matsmok+matalc+physfunc)))
```

The process is repeated for other elicited values of the sensitivity parameters shown in Table 2. As a check, we also set the sensitivity parameters to zero and check that results agree with the FCS analyses under MAR (using the same seed). The results are identical to the FCS results, which were 0.65 (95% CI: 0.22, 1.08).

The corresponding syntax for conducting NARFCS in Stata (using the "mi impute chained" command) is provided as Supplementary material. In brief, this requires the creation of a new variable that contains the values of the elicited sensitivity parameters for missing values only (e.g. create a new variable "offset_ matmhw1" that is equal to 0.11 if missing maternal mental illness, and is otherwise equal to 0). NARFCS is then implemented using the offset option in the "mi impute chained" command, e.g. "(logit, offset(offset_matmhw1))"

### *Step 5. Report results of sensitivity analysis*

The final step is to report the results of the NARFCS sensitivity analyses. As recommended by Lash and colleagues in the reporting of quantitative bias analyses,[51] the presentation of a sensitivity analysis should begin with a statement of the objectives of the sensitivity analyses. The methods used for the sensitivity analyses and any underlying assumptions should be described in detail. It should be stated whether sensitivity analyses were pre-planned (as recommended) or post-hoc, and results of all sensitivity analyses should be reported.[52]

Results from the NARFCS sensitivity analyses can be presented in tables or graphs. Figures may be preferred when tables are large and complex, and to enable patterns in results to be seen. In particular, graphs may be preferred when presenting results over a distribution of delta values, such as when using tipping point analyses. Tables may be preferred when there are 4 or more sensitivity parameters, as high-dimensional figures can be difficult to interpret.

When drawing overall conclusions, we suggest reporting the results of the sensitivity analysis alongside results obtained under the primary analysis, which as mentioned earlier is usually under MAR. In the study abstract, one can report results for a range of sensitivity values, and a statement regarding the extent to which results are robust to different assumptions concerning the missingness mechanism. If there is a distribution of delta values, the discussion of the results can be focused on results at the centre of the distribution, but providing results at the extreme ranges will shed light on the robustness of results under different assumptions. The sensitivity of the results to MNAR mechanisms should be interpreted in light of the magnitude of the effect sizes estimated in the sensitivity analysis and their likely clinical significance.

Returning to the LSAC case study, Figure 3 provides a heatmap showing the estimates of the target regression coefficient for each combination of the $\delta_Y$ and $\delta_X$ values shown in Table 2. The sensitivity analyses indicate that with increasing departures from MAR, the magnitude of the estimated regression coefficient generally increases. For example, looking at the results at the median value of $\delta_X$ (i.e. middle column of the heatmap), the estimated adjusted difference in mean SDQ scores between exposed and unexposed was 0.69 (95% confidence interval [CI]: 0.27, 1.10) when $\delta_Y$=-0.22, i.e. the $25^{th}$ percentile value. This increased to 0.72 (0.33, 1.10) at the median value of $\delta_Y$, and to 0.83 (0.41, 1.25) at the $75^{th}$ percentile value of $\delta_Y$. These results are consistent with those obtained under MAR (i.e. 0.65; 95% CI: 0.22 -1.08), particularly for results at the $25^{th}$ and $50^{th}$ values of $\delta_Y$ and $\delta_X$. Overall the conclusions remained the same across different values of the sensitivity parameters, i.e. there is a substantively small difference in SDQ scores when comparing those with mothers with and without probable mental illness (given that the range of SDQ scores is 0-40). Thus, for this sensitivity analysis, the study results were not sensitive to the changes we made to the missingness assumptions. Of note, when choosing a colour scale for a heatmap, researchers may consider the substantive meaning of the estimate values. For example, Figure 3 at first glance may suggest a big difference in estimates between bottom left and top right (very light to very dark), but substantively it is not a big difference.

[Figure 3 about here]

## DISCUSSION

In this tutorial, we have provided a roadmap for conducting sensitivity analysis for multivariable missing data using NARFCS. Using a case study, we illustrated the steps in this roadmap, including the use of m-DAGs to guide our decision-making for carrying out a sensitivity analysis. We also illustrated how to specify the delta-adjusted models and select values for the sensitivity parameters. Finally, we showed how to conduct NARFCS sensitivity analyses using standard FCS commands in R and Stata. It is hoped that the clear framework and syntax for carrying out sensitivity analyses provided will encourage the uptake of sensitivity analyses to departures from MAR in health research studies.

This tutorial focused on NARFCS as it is an accessible method that is now available in mainstream statistical software and builds on the widely used FCS MI approach. There are other approaches for carrying out sensitivity analysis that are not covered here. For example, in the context of randomised controlled trials, an alternative pattern-mixture approach is reference-based MI. This method does not require the explicit specification of sensitivity parameters; rather it imputes missing data assuming that those with missing values have similar behaviour to a reference group (e.g. the control arm).[37] Rather than using MI, it is also possible to fit pattern-mixture models within a fully Bayesian framework[39,53]. For example, Mason et al. elicited expert opinions regarding differences in quality of life measures for trial participants with and without missing values.[39] The elicited information was used to define informative priors in sensitivity analyses using a Bayesian analysis. Beyond pattern-mixture modelling, it is also possible to carry out sensitivity analysis using other frameworks such as selection modelling.[18] These approaches are beyond the scope of the current paper.

In this paper, we have illustrated NARFCS models where the univariate imputation models include only the missingness indicator corresponding to the variable to be imputed (e.g. in the LSAC example, the imputation model for the outcome included $M_Y$ only and the imputation model for the exposure included $M_X$ only). It would be possible to extend these NARFCS models to include missingness indicators for other variables as auxiliary variables. Tompsett et al. [23] recommend that NARFCS models should include the missingness indicators of other incomplete variables (e.g. $M_X$ in the model for imputing $Y$), as the associated regression coefficients can be estimated from the observed data, but they also acknowledge the lack of guidance available for the choice of missingness indicators to include in NARFCS models. The inclusion of missingness indicators in imputation models has been explored in the context of standard FCS [54]. In their simulation study, Beesley et al. [54] found that modifications to FCS (such as the inclusion of missingness indicators of other variables in the imputation models) led to reduced bias compared to standard FCS. However, their simulations did not include the MNAR setting where the probability of missingness in a variable depends on its own values. Outside of the FCS setting, Sperrin and Martin [55] also recommend the inclusion of missingness indicators (and their interactions) in imputation models, but with the caveat that the choice of indicators for a specific analysis requires careful consideration and should be guided by causal diagrams. Given the lack of guidance in this area, we have omitted missingness indicators of other variables in the NARFCS models in the case study (as is usually done with FCS), but this is an avenue of future research.

To be able to address all plausible problematic arrows or associations between variables and their missingness in practice, our general advice is to use a sensitivity parameter for each one (e.g. as for exposure and outcome in the LSAC example). There are, however, limits to how far one can go. If two variables are highly correlated, the combination of two positive sensitivity parameters may lead to an upward drift of the imputed values, possibly resulting in imputations far beyond the maxima in the data. A remedy for preventing such drift is to vary the sensitivity parameters one by one rather than simultaneously.

A limitation of this paper is that our illustrative example considered the simple setting of a point-exposure study, with a mix of binary and continuous variables. In practice, missing values may occur in other variable types (e.g. count data, ordinal data), for which NARFCS functions are not yet available in R. Furthermore, we may want to explore sensitivity analyses for missing data in more complex settings such as longitudinal analyses. For example, Ratitch et al. [20] and Moreno-Betancur and Chavance[25] illustrate the use of a delta-adjustment method within MI to handle longitudinal drop-out for randomised controlled trials. Further research is required to develop guidance on the application of NARFCS in settings with longitudinal and other multilevel data.

To date, the lack of flexible methods, software and guidance has likely been a barrier to the conduct of sensitivity analyses in real-world studies, particularly in the context of multivariable missingness. Our tutorial aimed at addressing this barrier by providing a step-by-step guidance into the use of NARFCS, a practical and accessible approach for performing sensitivity analyses for multivariable missing data. The recent accessibility of NARFCS in R and Stata will hopefully improve the uptake of sensitivity analyses to the missing data mechanism, which to date are still infrequently performed despite recommendations.[17]

**FUNDING**

This work was supported by a grant from the Australian National Health and Medical Research Council: Project Grant ID 1166023, Career Development Fellowship to KJL (ID 1127984) and Investigator Grant to MMB (ID 2009572). IW was supported by the Medical Research Council Programme MC_UU_00004/07. Research at the Murdoch Children's Research Institute is supported by the Victorian Government's Operational Infrastructure Support Program. The funding bodies do not have any role in the collection, analysis, interpretation or writing of the study.

**ACKNOWLEDGEMENTS**

This paper uses unit record data from Growing Up in Australia: the Longitudinal Study of Australian Children conducted by the Australian Government Department of Social Services (DSS). The findings and views reported in this paper, however, are those of the authors and should not be attributed to the Australian Government, DSS or any of DSS's contractors or partners. DOI: 10.26193/JOZW2U.

**CONFLICTS OF INTEREST**

None

*Table 1. Summary of variables relevant to the motivating example (n=4882)*

| Variable description | Variable label in dataset | Variable type | Notation | LSAC wave | Range/values | Missing data n(%) |
|---|---|---|---|---|---|---|
| Sex of the child | sex | Confounder | $Z_1$ | 1 | 1 – Female <br> 0 – Male | 0 (0%) |
| Child has a sibling | siblings | Confounder | | 1 | 1 – Yes <br> 0 – No | 0 (0%) |
| Mother completed high school | matedu | Confounder | | 1 | 1 – Yes <br> 0 – No | 0 (0%) |
| Maternal age | matage | Confounder | | 1 | Years | 0 (0%) |
| Consistent parenting | conspar | Confounder | | 1 | Range 1-5 | 0 (0%) |
| Financial hardship | finhard | Confounder | | 1 | 1 – Yes <br> 0 – No | 0 (0%) |
| SDQ score | basesdq | Confounder | | 1 | Range 0-40 | 0 (0%) |
| Maternal smoking status | matsmok | Confounder | $Z_{21}$ | 1 | 1 – Yes <br> 0 – No | 771 (16%) |
| Maternal risky alcohol drinking | matalc | Confounder | $Z_{22}$ | 1 | 1 – Yes <br> 0 – No | 885 (18%) |
| Child's physical functioning score | physfunc | Confounder | $Z_{23}$ | 1 | Range 0-100 | 742 (15%) |
| Maternal mental illness (Kessler<4) | matmhw1 | Exposure | $X$ | 1 | 1 – Probable serious mental illness <br> 0 – No mental illness | 738 (15%) |
| Child behaviour (SDQ score) | sdqw3 | Outcome | $Y$ | 3 | Range 0-40 | 1142 (23%) |

*Table 2. Values of the sensitivity parameters used in the LSAC case study analysis (based on values elicited by Hayati Rezvan et al. 2018)*

|  |  | $\delta_X$ | | |
|---|---|---|---|---|
|  |  | 25th percentile MSP$_X$=0.2 | 50th percentile MSP$_X$=0.4 | 75th percentile MSP$_X$=0.7 |
| $\delta_Y$ | 75th percentile MSP$_Y$=5 | $\delta_Y$=3.8, $\delta_X$=-0.19 | $\delta_Y$=3.8, $\delta_X$=0.08 | $\delta_Y$=3.8, $\delta_X$=0.43 |
|  | 50th percentile MSP$_Y$=2 | $\delta_Y$=0.79, $\delta_X$=-0.15 | $\delta_Y$=0.78, $\delta_X$=0.11 | $\delta_Y$=0.77, $\delta_X$=0.47 |
|  | 25th percentile MSP$_Y$=1 | $\delta_Y$=-0.21, $\delta_X$=-0.13 | $\delta_Y$=-0.22, $\delta_X$=0.13 | $\delta_Y$=-0.23, $\delta_X$=0.48 |

Note: $\delta_Y$=conditional sensitivity parameter for *Y*, i.e. coefficient of the missingness indicator in the delta-adjusted model for *Y*; $\delta_X$=conditional sensitivity parameter for *X*, i.e. coefficient of the missingness indicator in the delta-adjusted model for *X*. MSP$_Y$=marginal sensitivity parameter for *Y*; MSP$_X$=marginal sensitivity parameter for *X*. The values for MSP$_Y$ and MSP$_X$ were the 25th, 50th and 75th percentiles of the pooled expert-elicited distributions for the sensitivity parameters reported in Hayati Rezvan et al. (2018), with rounding to one significant figure. The values of the conditional sensitivity parameters $\delta_Y$ and $\delta_X$ were obtained by jointly calibrating the MSP values using methods described in Tompsett et al. (2018).

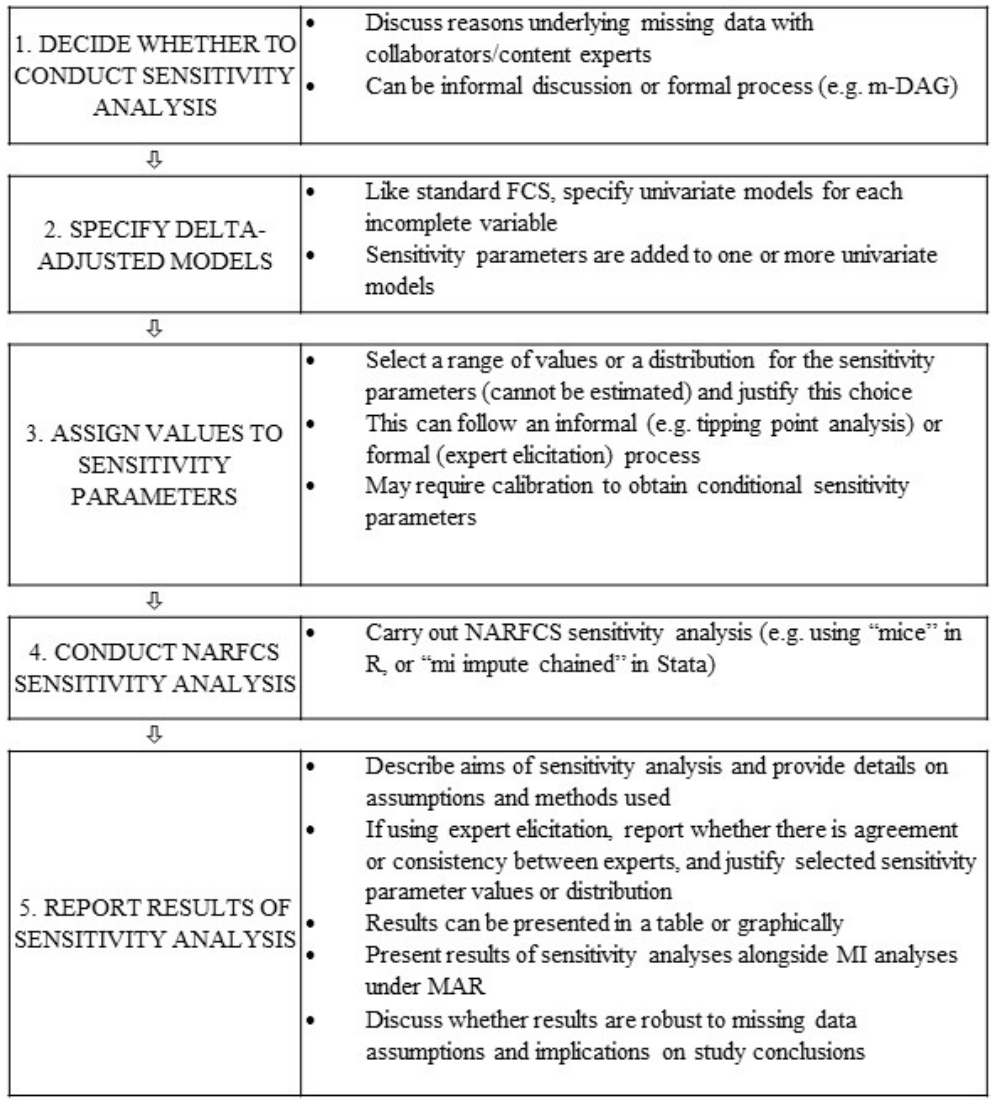

*Figure 1. Schematic of a roadmap for conducting a NARFCS sensitivity analysis*

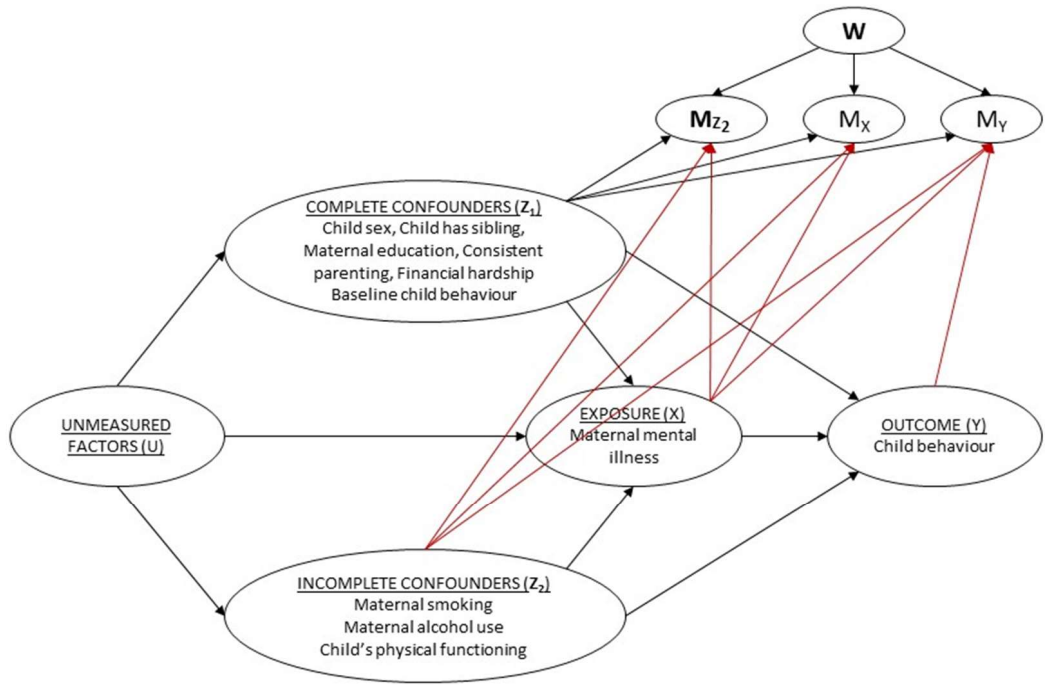

*Figure 2. Missingness-DAG for the case study investigating the effect of maternal mental illness on child behaviour using data from the Longitudinal Study of Australian Children. Adapted from Figure 1 of Moreno-Betancur et al.[25]*

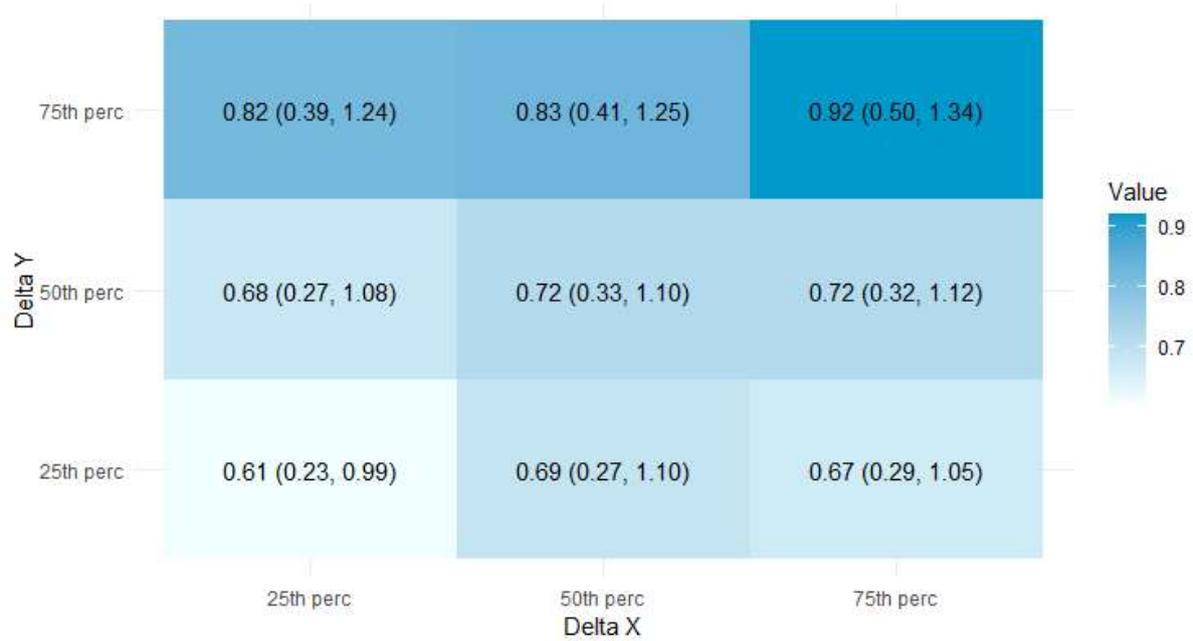

*Figure 3. Heatmap of estimates of the regression-adjusted difference in SDQ scores (with 95% CI) between children with and without mothers with probable serious mental illness at different combinations of $\delta_Y$ and $\delta_X$, corresponding to the sensitivity parameter values shown in Table 2.*